\documentclass[letterpaper, 10 pt, conference]{ieeeconf}

\IEEEoverridecommandlockouts                              

\usepackage{tikz}
\usepackage{textcomp}
\usepackage{hyperref}
\usepackage{lipsum}

\newcommand\copyrighttext{%
  \footnotesize \textcopyright 2020 IEEE.  Personal use of this material
    is permitted. Permission from IEEE must be obtained for all other uses, in
    any current or future media, including reprinting/republishing this material
    for advertising or promotional purposes, creating new collective works, for
    resale or redistribution to servers or lists, or reuse of any copyrighted
    component of this work in other works}
\newcommand\copyrightnotice{%
\begin{tikzpicture}[remember picture,overlay]
\node[anchor=south,yshift=10pt] at (current page.south) {\fbox{\parbox{\dimexpr\textwidth-\fboxsep-\fboxrule\relax}{\copyrighttext}}};
\end{tikzpicture}%
}
\usepackage{graphics} 
\graphicspath{{./figures/}}
\usepackage{epsfig} 
\usepackage{booktabs}
\usepackage{tabularx}

\usepackage{url}

\title{\LARGE \bf BigO: A public health decision support system for measuring obesogenic
  behaviors of children in relation to their local environment}

\author{Christos Diou$^{1}$, Ioannis Sarafis$^{1}$, Vasileios Papapanagiotou$^{1}$, Leonidas Alagialoglou$^{1}$, Irini Lekka$^{1}$, \\
  Dimitrios Filos$^{1}$, Leandros Stefanopoulos$^{1}$, Vasileios Kilintzis$^{1}$, Christos Maramis$^{1}$, Youla Karavidopoulou$^{1}$, \\
  Nikos Maglaveras,$^{1}$, Ioannis Ioakimidis$^{2}$, Evangelia Charmandari$^{3}$, Penio Kassari$^{3}$, Athanasia Tragomalou$^{3}$,\\
  Monica Mars$^{4}$, Thien-An Ngoc Nguyen$^{5}$, Tahar Kechadi$^{5}$, Shane O’Donnell$^{5}$, Gerardine Doyle$^{5,6}$,\\
  Sarah Browne$^{7}$, Grace O' Malley$^{7,8}$, Rachel Heimeier$^{9}$, Katerina
  Riviou$^{10}$, Evangelia Koukoula$^{11}$, \\
  Konstantinos Filis$^{12}$, Maria Hassapidou$^{13}$, Ioannis Pagkalos$^{13}$,
  Daniel Ferri$^{14}$, Isabel P\'{e}rez$^{14}$\\
  and Anastasios Delopoulos$^{1}$\\
\thanks{*The work leading to these results has received funding from
    the European Community's Health, demographic change and well-being
    Programme under Grant Agreement No. 727688, 01/12/2016 -
    30/11/2020 (http://bigoprogram.eu/).}
\thanks{$^{1}$Aristotle University of Thessaloniki, Greece}%
\thanks{$^{2}$Karolinska Institutet, Sweden}%
\thanks{$^{3}$Biomedical Research Foundation of the Academy of Athens, Greece}%
\thanks{$^{4}$Wageningen University, Netherlands}%
\thanks{$^{5}$University College Dublin, Ireland}%
\thanks{$^{6}$UCD Geary Institute for Public Policy, Ireland}%
\thanks{$^{7}$Childrens' Health Ireland at Temple Street, Ireland}%
\thanks{$^{8}$Royal College of Surgeons, Dublin, Ireland}%
\thanks{$^{9}$International English School, Sweden}%
\thanks{$^{10}$Ellinogermaniki Agogi, Greece}%
\thanks{$^{11}$Ekpaideftiria N. Mpakogianni, Greece}%
\thanks{$^{12}$COSMOTE, Greece}%
\thanks{$^{13}$International Hellenic University, Greece}%
\thanks{$^{14}$Mysphera, Spain}%
}

\begin{document}
\bstctlcite{IEEEexample:BSTcontrol}

\maketitle
\thispagestyle{empty}
\pagestyle{empty}
\copyrightnotice

\begin{abstract}
Obesity is a complex disease and its prevalence depends on multiple factors related to the
local socioeconomic, cultural and urban context of individuals. Many obesity prevention
strategies and policies, however, are horizontal measures that do not depend on
context-specific evidence. In this paper we present an overview of BigO
(\url{http://bigoprogram.eu}), a system designed to collect objective behavioral data from
children and adolescent populations as well as their environment in order to support
public health authorities in formulating effective, context-specific policies and
interventions addressing childhood obesity. We present an overview of the data
acquisition, indicator extraction, data exploration and analysis components of the BigO
system, as well as an account of its preliminary pilot application in 33 schools and 2
clinics in four European countries, involving over 4,200 participants.
\end{abstract}

\section{INTRODUCTION}
\label{sec:introduction}
Obesity prevalence has been continuously rising for the past forty years
\cite{euac2014} and is now one of the world's biggest health challenges. Given
that the disease is largely preventable, researchers have sought appropriate
policy measures to limit the development of overweight and obesity, especially
in children, since children who are overweight or obese are likely to remain
obese in adulthood \cite{lobstein2015}.

Many large-scale public health actions are limited to indiscriminate blanket policies and
single-element strategies \cite{oude2009}, that often fail to address the problem
effectively. This has been attributed to the complex nature of the disease
\cite{finegood2010}, implying that effective interventions must be evidence-based, adapted
to the local context and address multiple obesogenic factors of the environment
\cite{oude2009, mackenbach2014}, even on a local neighborhood level \cite{Fagerberg2019}.

Developing effective multi-level interventions addressing childhood obesity therefore
requires data that link conditions in the local environment to children's obesogenic
behaviors such as low levels of physical activity, unhealthy eating habits, as well as
insufficient sleep.

Most of the existing evidence on obesogenic behaviors of
children rely on diet and physical activity recall questionnaires
\cite{richardson2001}, which can lead to inaccurate measurements
\cite{dhurandhar2015} and often do not provide sufficient detail about the
interaction of children with their environment (such as use of available
opportunities for physical activity, or visits to different types of food
outlets).

On the other hand, the availability and widespread use of wearable and portable
devices, such as smartphones and smartwatches, provide an excellent opportunity
for obtaining objective measurements of population behavior. This has not yet
been exploited to its full extent for evidence-based policy decision support
addressing childhood obesity.
\begin{figure*}[ht]
  \centering
  \includegraphics[width=0.95\linewidth]{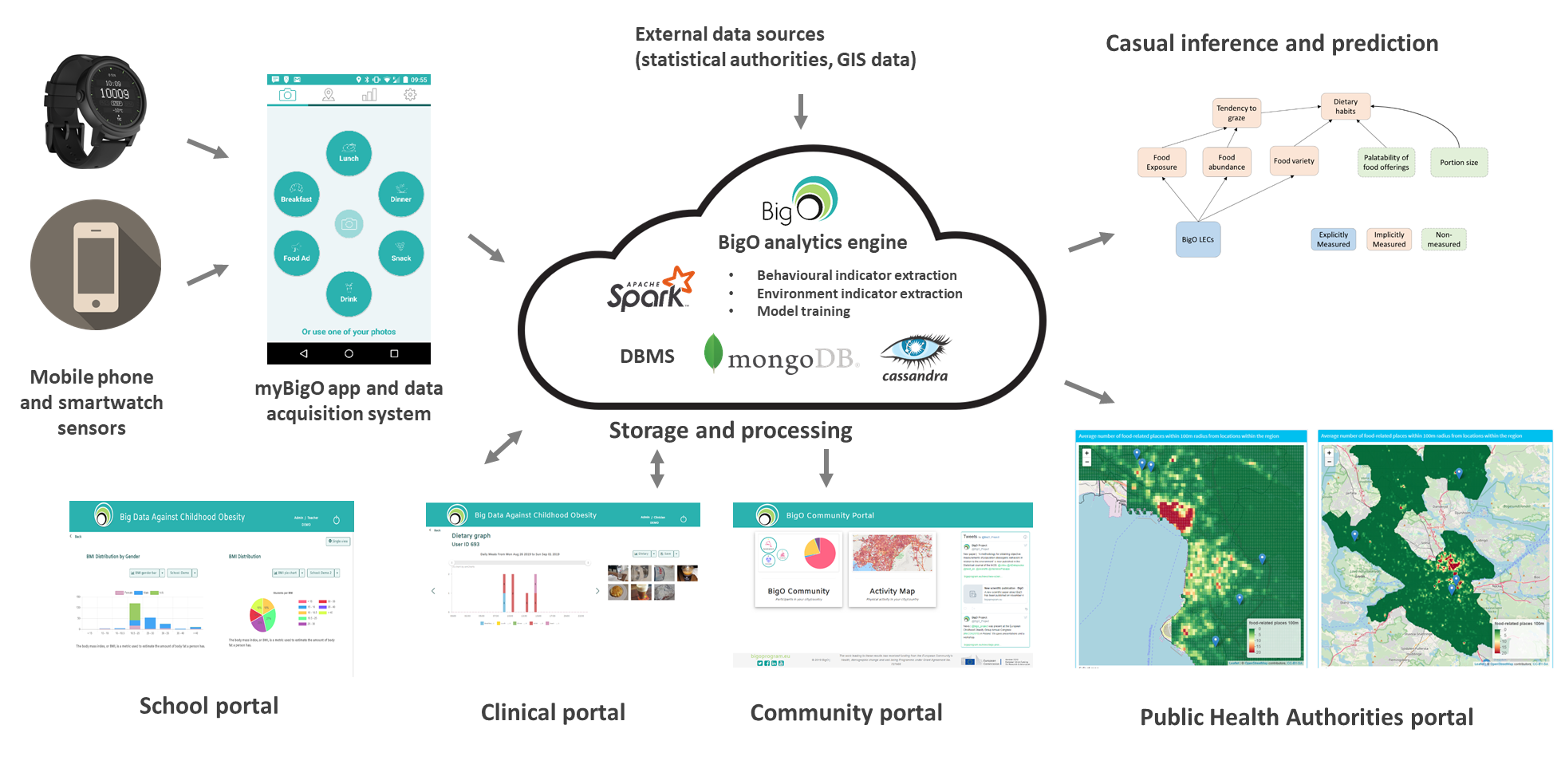}
  \caption{BigO system overview. Data is collected by children and their
    environment using mobile and smartwatch applications as well as other online
    data sources. The data is processed to extract individual and aggregated
    behavior and environment indicators. Aggregated indicators are then used
    through the Public Health Portal, a web application that supports data
    exploration and visualization and helps analysts make inferences about
    possible local drivers of obesogenic behaviors, as well as to assess and
    predict the impact of policy interventions addressing childhood
    obesity. Other web applications (portals) are also provided. The School
    portal is used to organize educational activities around obesity at school
    and to coordinate class or group participation in the data collection. The
    Clinical portal is used by clinicians to better measure their patients'
    obesogenic behaviors and provide personalized guidance. Finally, the
    Community portal provides a summary of data and findings to the public.}
  \label{fig:overview}
\end{figure*}

In this paper we present an overview of the BigO
system\footnote{\url{http://bigoprogram.eu}}, which has been developed with the
aim to objectively measure the obesogenic behaviors of children and adolescent
populations in relation to the local environment using a smartphone and
smartwatch application. Specifically, BigO provides policy makers the tools to
(i) measure the behavior of a sample of the local population (targeting ages
9-18) regarding their physical activity, eating and sleep patterns, (ii)
aggregate data at geographical level to avoid revealing any individual
information about participants, (iii) quantify the conditions of the local urban
environment, (iv) visualize and explore the collected data, (v) perform
inferences about the strength of relationships between the environment and
obesogenic behaviors and, finally, (vi) predict and monitor the impact of policy
interventions addressing childhood obesity.

The following sections present an outline of the BigO system as well as the
challenges that we have identified from its preliminary application to data
collection pilots in schools and clinics, involving over 4,200 participants to
date (April 2020).

\section{SYSTEM OVERVIEW}
\label{sec:overview}

A conceptual overview of the BigO system is shown in Figure
\ref{fig:overview}. Data is collected from smartwatch and smartphone sensors
through the ``myBigO'' app, available for Android and iOS operating systems
\cite{maramis2019}. It is then stored at the BigO DBMS, consisting of Apache
Cassandra (for time series data) and MongoDB (for all other application data)
databases. Data processing is carried out by the BigO analytics engine, built on
top of Apache Spark.

The processing steps involve the extraction of individual and population-level
behavioral indicators, the extraction of environment indicators, as well as
statistical data analysis and machine learning model training. The processing
outputs support the operation of the end-user interfaces which include the
Public Health Authorities portal, School portal, Clinical Portal and the
Community portal.

\subsection{Data acquisition}
\label{sec:data_acquisition}

Administrative and operational data (e.g., number of exercise sessions per week
at school, availability of school lunches, class start/end times) are collected
at school and clinic level through the portals. Furthermore, data about body
mass index range, age and sex of participants is also collected through the
portal. No directly identifiable information (such as names or emails) is stored
anywhere in the system.

Regarding individual participants, data is collected through a smartphone and
smartwatch application. Collected data includes (i) triaxial accelerometer
signal, (ii) GPS location data, (iii) meal pictures, (iv) food advertisement
pictures, (v) meal self-reported data, (vi) answers to a one-time questionnaire
(when the myBigO app is started for the first time) and (v) answers to recurring
mood questionnaires.

The biggest challenges in data collection come from the battery power
requirements of the accelerometer and location sensors. Accelerometer is sampled
at a low sampling rate, which is device dependent and is usually in the range of
5-15Hz. Location data is sampled every minute. To preserve battery, the data
acquisition module of the mobile application is compatible with the ``doze''
mode of the Android operating system. It stops data capturing whenever a device
is inactive (doze mode) and restarts whenever the device becomes active
again. During this time, acceleration is assumed to only be affected by gravity
(not any kind of user motion) and location data is fixed to the last known position
\cite{diou2019JIAOS}

\subsection{Extraction of behavioral and environment indicators}
\label{sec:indicators}

Collected raw data are processed to extract behavioral indicators. This can take
place at the mobile phone (to avoid transmitting raw data) or centrally, at the
BigO servers. In both cases, raw data is considered sensitive and cannot be
accessed directly. Aggregated behavioral indicators that result from the raw
data are used instead \cite{diou2019JIAOS}. There are three levels of
granularity of behavioral indicators, namely (i) base indicators, which describe
the behavior of an individual at fine temporal granularity (ii) individual
indicators, which aggregate indicators across time to summarize the behavior of
an individual and (iii) population indicators, which aggregate the behavior
across individuals in a geographical region. Examples are provided in Tables
\ref{tab:base_indicators} and \ref{tab:indicators_lecs}.

Base indicators are extracted through signal processing and machine learning
algorithms \cite{papapanagiotou2020a}, such as \cite{gu2017robust} and
\cite{genovese2017smartwatch} for step counting, \cite{reiss2012creating} for
activity type detection, \cite{shafique2016travelo} for transportation mode
detection and \cite{luo2017improved} for detecting visited points of interest
(POIs). Regarding the visited POIs, the information stored is the type of POI,
from a pre-defined POI hierarchy, such as ``restaurant'', ``fast food or
takeaway'' and ``sports facility''.
\begin{table}
  \centering
  \caption{Base behavior indicators used in BigO}
  \label{tab:base_indicators}
  \begin{tabular}{ll}
    \toprule
    \textbf{Type of behavior} & \textbf{Base behavior indicator}\\
    \midrule
    \emph{Physical activity} & Activity counts\\
    & Number of steps \\
    & Physical activity type \\
    \emph{Location and transportation} & Types of visited locations \\
    & Transportation mode used \\
    \emph{Eating habits} & Visits to food-related locations\\
    & Meals (self-reported, with pictures) \\
    \emph{Sleep} & Sleep duration (only for smartwatch users)\\
    \bottomrule
  \end{tabular}
\end{table}

Certain base indicators (such as activity counts) and individual indicators are
only used at the clinical portal, which is accessible by health professionals to
obtain information about their patients. In all other cases, aggregated
information is used, for privacy protection purposes. There are two types of
aggregation, leading to two different analysis types:
\begin{itemize}
  \item \emph{Habits}: In this analysis we are interested in the overall behavior of participants
    \emph{living} in a region
  \item \emph{Use of resources}: In this analysis we are interested in the behavior
      of participants \emph{visiting} a region, but only during their visits to that
      region
\end{itemize}
A region can either be an administrative region or a geohash\footnote{A geohash
  is a public domain geocode system encoding rectangular geographical regions as
  alphanumeric strings}.

In addition to the measured behavior of individuals, each geographical region is
characterized by the local urban and socio-economic context. These are
quantified by a set of variables called \emph{Local Extrinsic Conditions} (LECs)
in BigO, which are obtained through official statistics, or through GIS
databases. Table \ref{tab:indicators_lecs} shows some examples of LECs used in BigO.
\begin{table}
  \centering
  \caption{Examples of population behavior indicators and LECs}
  \label{tab:indicators_lecs}
  \begin{tabularx}{.45\textwidth}{p{2.3cm}X}
    \toprule
    \textbf{Variable} & \textbf{How it is computed}\\
    \midrule
    \multicolumn{2}{l}{\emph{Examples of behavior indicators}}\\
    Percentage of visits to the region that include at least one visit to a fast
    food or a takeaway restaurant &
    Compute the percentage of visits to the
    region that include at least one visit to a fast food or takeaway
    restaurant. For the computation of the indicator, only visits that are 10
    minutes or more are considered.\\
    & \\
    Percentage of residents that have an average activity level that is
    sedentary &
    For each resident of the region, first compute their number of
    steps for each minute of recorded data, for individuals with more than 20
    hours of recorded data. This indicator is the percentage of residents of the region
    that walk, on average, less than 450 steps per hour. \\
    \midrule
    \multicolumn{2}{l}{\emph{Examples of Local Extrinsic Conditions}}\\
    Average number of restaurants within 100m radius from locations within the region &
    Create a 30m point grid inside the region. For each point, compute the number
    of restaurants in a 100m radius. This value is the average across all points
    inside the region.\\
    & \\
    Number of athletics/sports facilities in the region & Using publicly
    available data sources compute the number of athletics/sports facilities
    inside the region. \\
    \bottomrule
  \end{tabularx}
\end{table}

\subsection{Data analysis for causal inference and prediction}
\label{sec:data_analysis}

The collected behavior and environment indicators can be used to (i) infer associations
and possible causal relations between LECs and obesogenic behaviors and (ii) predict and
monitor the impact of interventions on the measured population behaviors.

\begin{figure}
  \centering
  \includegraphics[width=0.98\linewidth]{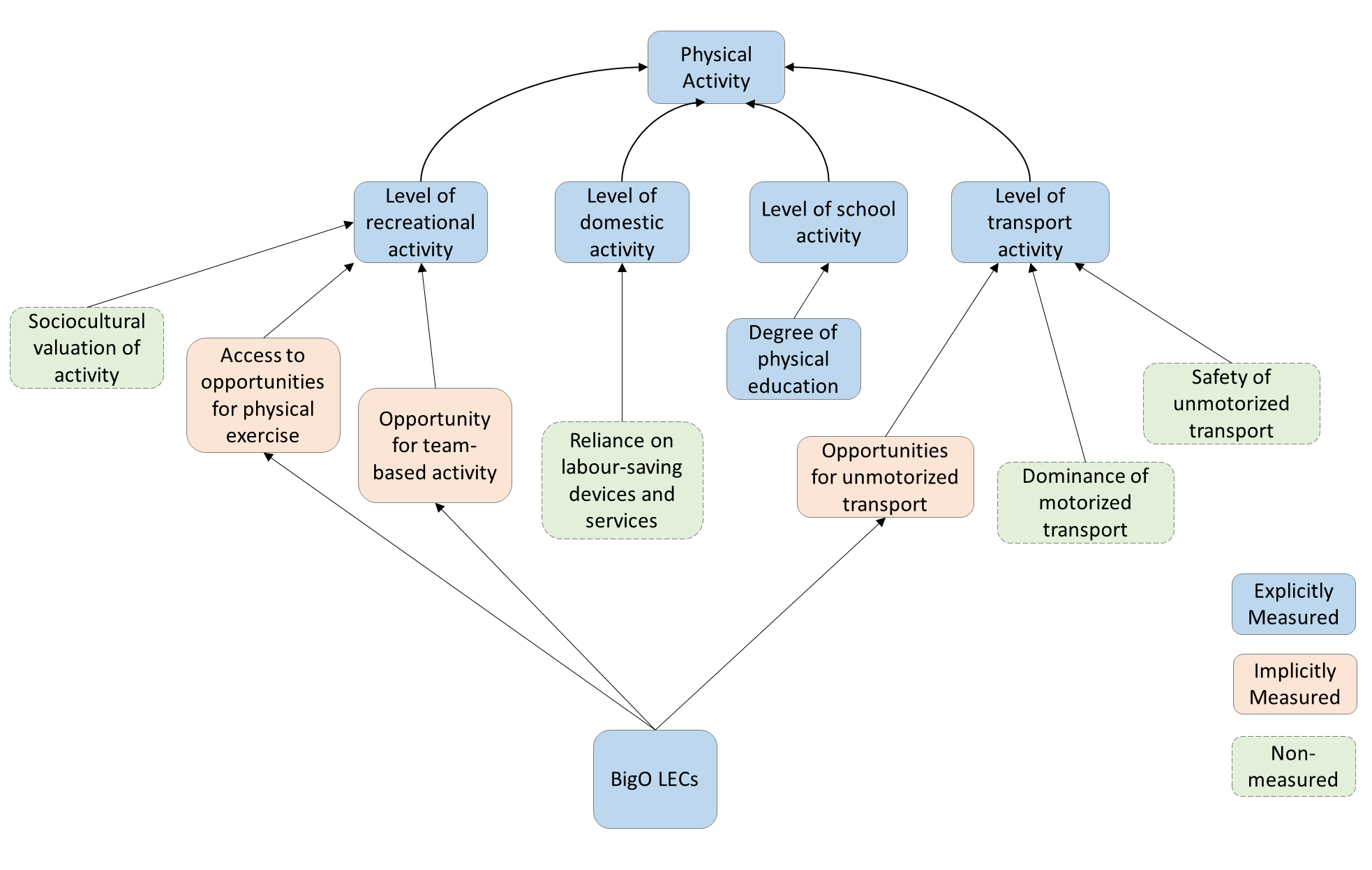}
  \caption{Causal model for physical activity. Physical activity is measured directly
    through behavior indicators. Some of the mediators are only implicitly measured. }
  \label{fig:dag}
\end{figure}

Given the complexity and multifactorial nature of obesity
\cite{butland2007tackling}, it is important to account for confounding factors
and to avoid spurious correlations. To this end, we start from the Foresight's
Obesity System Map \cite{butland2007tackling} and construct Directed Acyclic
Graphs (DAGs) indicating causal relations between variables. An example DAG for
physical activity is shown in Figure \ref{fig:dag}.

The associations between variables are quantified using explainable statistical
models, such as linear and generalized linear models. Self-selection and other
sources of bias will need to be quantified during the analysis as well. For
prediction, non-linear machine learning models can be used as well, such as
Support Vector Machines, or Neural Networks. The analysis of BigO data is still
work in progress, however some preliminary results on prediction are presented
in \cite{sarafis2020}.

\section{DATA COLLECTION PILOTS}
\label{sec:pilots}
BigO has been deployed in clinics and in schools in Athens, Larissa and
Thessaloniki in Greece, the Stockholm area in Sweden and Dublin in
Ireland. Children join as citizen-scientists and contribute data about their
behavior and environment through the myBigO app. The planned data collection
time is two weeks per child. In schools, data is used in school projects, with
the help of the BigO School portal, which provides data visualizations for
participating school classes. In the clinic, the data is used by clinicians to
monitor the behavior of patients, through the Clinical portal.

In total, children from 33 schools and 2 clinics have contributed data so
far. Ethical approvals have been obtained, as well as the necessary consent from
all participants. By April of 2020, BigO had reached out to 20,000 children and
their parents, out of which over 4,200 registered in the system. Not all
children provide the same amount of data. Reasons for children providing fewer
data than expected include technical issues from the user side (e.g. smartwatch
not properly paired with phone), technical issues with the smartphone (e.g.,
some smartphone manufacturers do not permit background apps to run) as well as
low participant compliance. The current estimate is that monitoring data is
received for approximately 68\% of the app usage time, while approx 25\% of the
registered users do not provide accelerometer or GPS data (only self-reports and
pictures). The currently collected data volume includes approximately 107 years
of accelerometery data, 73 years of GPS data and 75,000 meal pictures. Note that
the actual monitoring time is higher (since no data is recorded when the device
is idle).

\section{DISCUSSION AND CONCLUSIONS}
\label{sec:conclusions}

There are some noteworthy observations that result from the experience in
organizing and deploying the BigO pilots.

On the technical side, there are significant obstacles to overcome when
depending on off-the-shelf devices such as smartphones and smartwatches for data
acquisition. Besides battery consumption, certain mobile phone manufacturers
introduce custom modifications to the device operating system which can prevent
background recording applications to run. Users must be aware of these
restrictions and disable them for the myBigO app, a process which is
device-dependent and not always straightforward.

Regarding recruitment, it seems that engaging teachers and clinicians is an
effective way to invite the participation of children and their parents (who
must give their consent). So far, the BigO portals have been used by 68 teachers
and 17 clinicians, leading to an acceptance rate of approximately 21\% for the
children that were reached out to participate in BigO. This approach is now
challenged by the recent school lockdowns due to the SARS-CoV-2 pandemic, but
recruitment is expected to resume once schools open again.

It is clear that scaling up such data collection actions requires the active
engagement of the local school and education authorities. In BigO, the pilots
were carried through the initiative of participating researchers and
schools/clinics that decided to join, without the direct support from the local
authorities. Our vision is that by demonstrating that such citizen-science
activities are effective for collecting data to formulate evidence-based
policies, BigO will motivate local authorities to adopt such systems as part of
their decision-making process. While data collection in the BigO pilots
continues, the next steps include the analysis of the collected data and the
dissemination of results to all relevant regional and national government
bodies.


\bibliographystyle{IEEEtran}
\bibliography{biblio}

\end{document}